\documentclass[a4paper]{article}

\usepackage{icrc2013}
\usepackage[english]{babel}
\usepackage{epstopdf}

\newcommand{\rarr}{\rightarrow}

\newcommand{\fbar}{{\bar f}}
\newcommand{\pbar}{{\bar p}}
\newcommand{\Psibar}{{\bar\Psi}}

\newcommand{\ubar}{{\overline u}}

\mathchardef\mhyphen="2D 
\newcommand{\beq}[1]{\begin{equation}\label{#1}}
\newcommand{\eeq}{\end{equation}}
\newcommand{\bea}[1]{\begin{eqnarray}\label{#1}}
\newcommand{\eea}{\end{eqnarray}}

\newcommand{\rf}[1]{(\ref{#1})}

\newcommand{\barr}{\begin{array}}
\newcommand{\earr}{\end{array}}
\def\be{\begin{equation}}
\def\ee{\end{equation}}
\def\ba{\begin{eqnarray}}
\def\ea{\end{eqnarray}}
\keywords{icrc2013, Dark Majorana annihilation, indirect Dark Matter detection, electroweak bremsstrahlung.}

\title{Likely dominance of WIMP annihilation to fermion pair\ +W/Z}
\shorttitle{Majorana WIMP annihilation to fermion pair+W/Z}

\authors{
Thomas J. Weiler
}

\afiliations{
Department of Physics \& Astronomy, Vanderbilt University, Nashville TN 37235 USA
}

\email{tom.weiler@vanderbilt.edu}

\abstract{
Arguably, the most popular candidate for Dark Matter (DM) is a massive, stable, Majorana fermion.  
However, annihilation of Majorana DM to two fermions features a helicity-suppressed $s$-wave rate.
The process radiating a gauge boson via electroweak (EW) and electromagnetic (EM)~bremsstrahlung 
removes this $s$-wave suppression, and is 
likely to be the dominant modes of gauge-singlet Majorana DM annihilation.
Given their enhanced annihilation rate with radiated $W$ and $Z$ gauge bosons, and the subsequent 
dominant $W/Z$ ~decays via hadronic channels, Majorana DM  tends to produce more antiprotons than positrons.
This result contrasts with observations, thereby presenting a challenge to model building with Majorana DM.
}

\begin{document}

\maketitle

\section{Introduction}
\label{sec:storyline}
The ``WIMP miracle'' (WIMP, for Weakly Interacting Massive Particle) for Dark Matter (DM) physics is that 
$\sigma_{\rm ann}\sim (\frac{\alpha_{\rm weak}}{4\pi})^2/M_{\rm WIMP}^2 \sim$~pb$\times(\frac{200\,{\rm GeV}}{M_{\rm WIMP}})^2$
yields the critical thermally-averaged annihilation ``rate''
$\langle{\rm v}\sigma\rangle \sim 3\times 10^{-26}{\rm cm}^3/s$
to effect a DM abundance in agreement with what is observed today, $\Omega_{\rm DM}\sim 0.24$.
Here, v~is the DM M\"oller velocity.
% ~\cite{note1}
% in the CoM frame
% and $\sigma_{\rm ann}$ is the DM annihilation cross section.
Among the many candidates for DM, the most common particle-type is the Majorana fermion  ($\chi$).
A splendid example is the neutralino  of supersymmetry models;
being a partner to a massless gauge-boson in the unbroken SUSY theory, the neutralino has two independent spin states, not four.  
Thus, $\chi$ is a Majorana fermion.
Here we focus not on iso-vector DM like the neutralino, but rather on gauge-singlet Majorana DM.
(Note that for gauge-singlet DM, 
the decay to $W^+\,W^-$ and $ZZ$ modes is very suppressed by quantum number conservation.)

\begin{figure*}[t]
%\begin{figure}[t]
%\centering
\includegraphics[height=0.10\textheight,width=0.32\columnwidth]{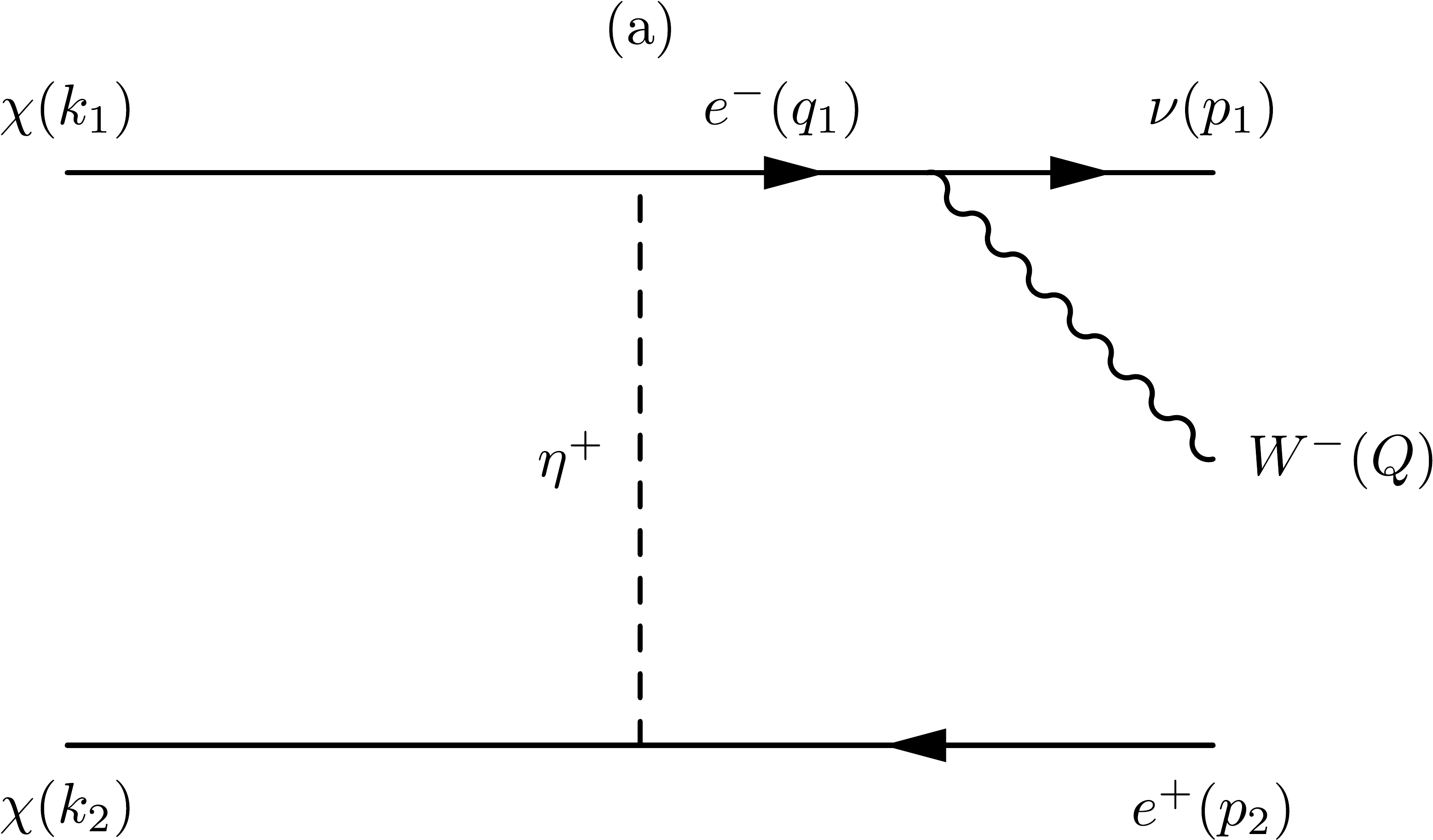}
$\;$
\includegraphics[height=0.10\textheight,width=0.32\columnwidth]{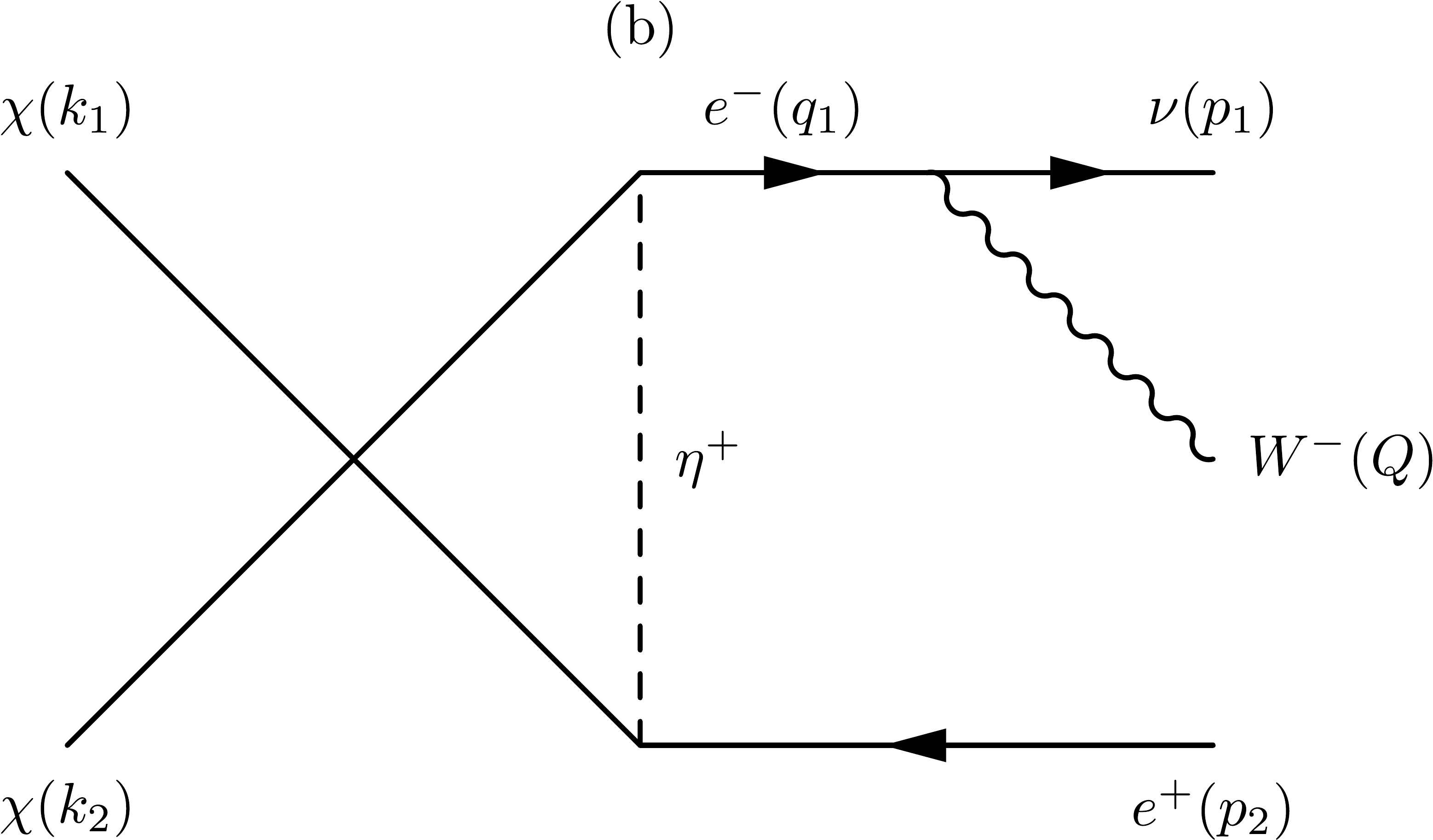}
$\;$
\includegraphics[height=0.10\textheight,width=0.32\columnwidth]{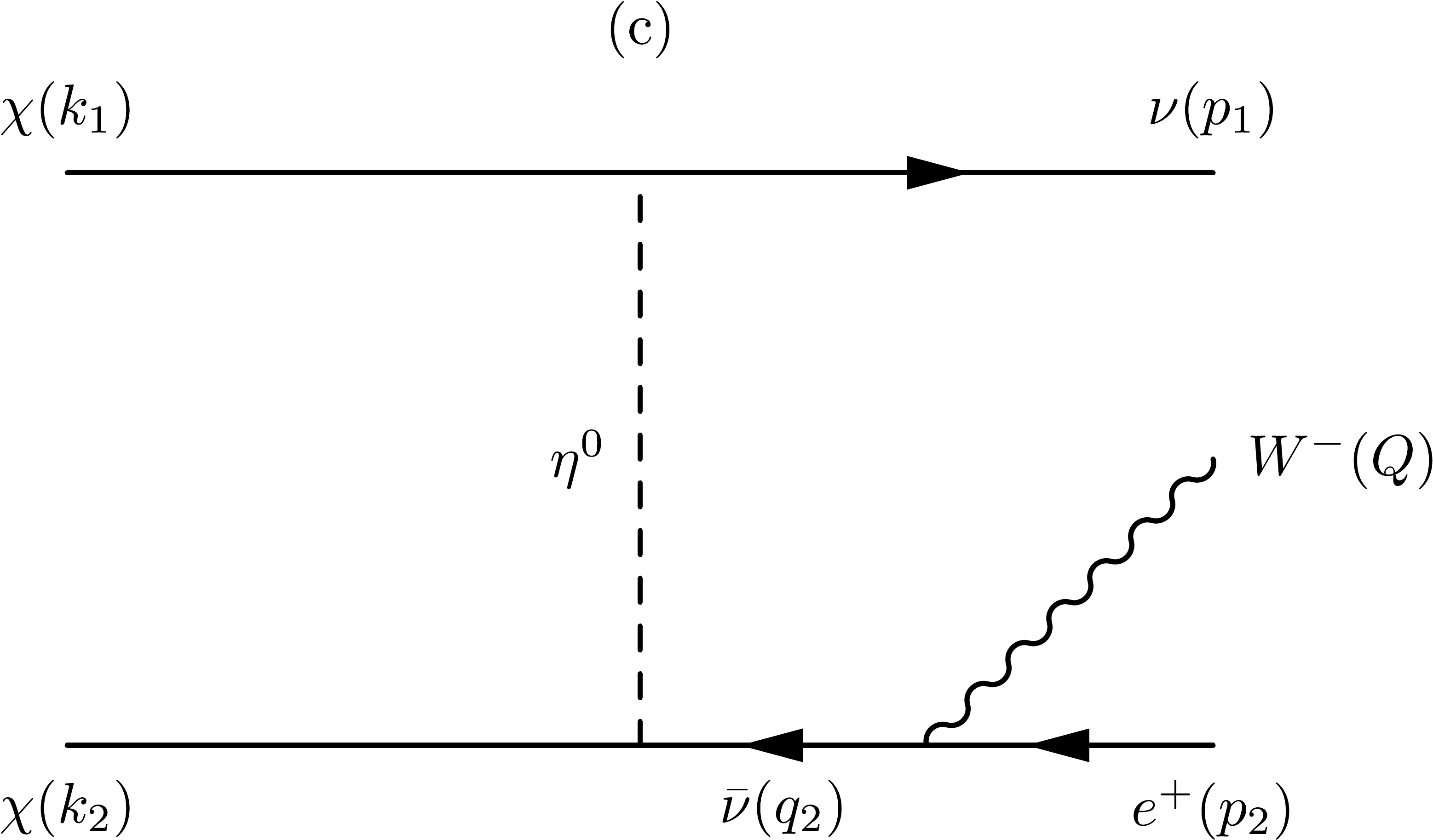}
$\;\;\;\;$
\includegraphics[height=0.10\textheight,width=0.32\columnwidth]{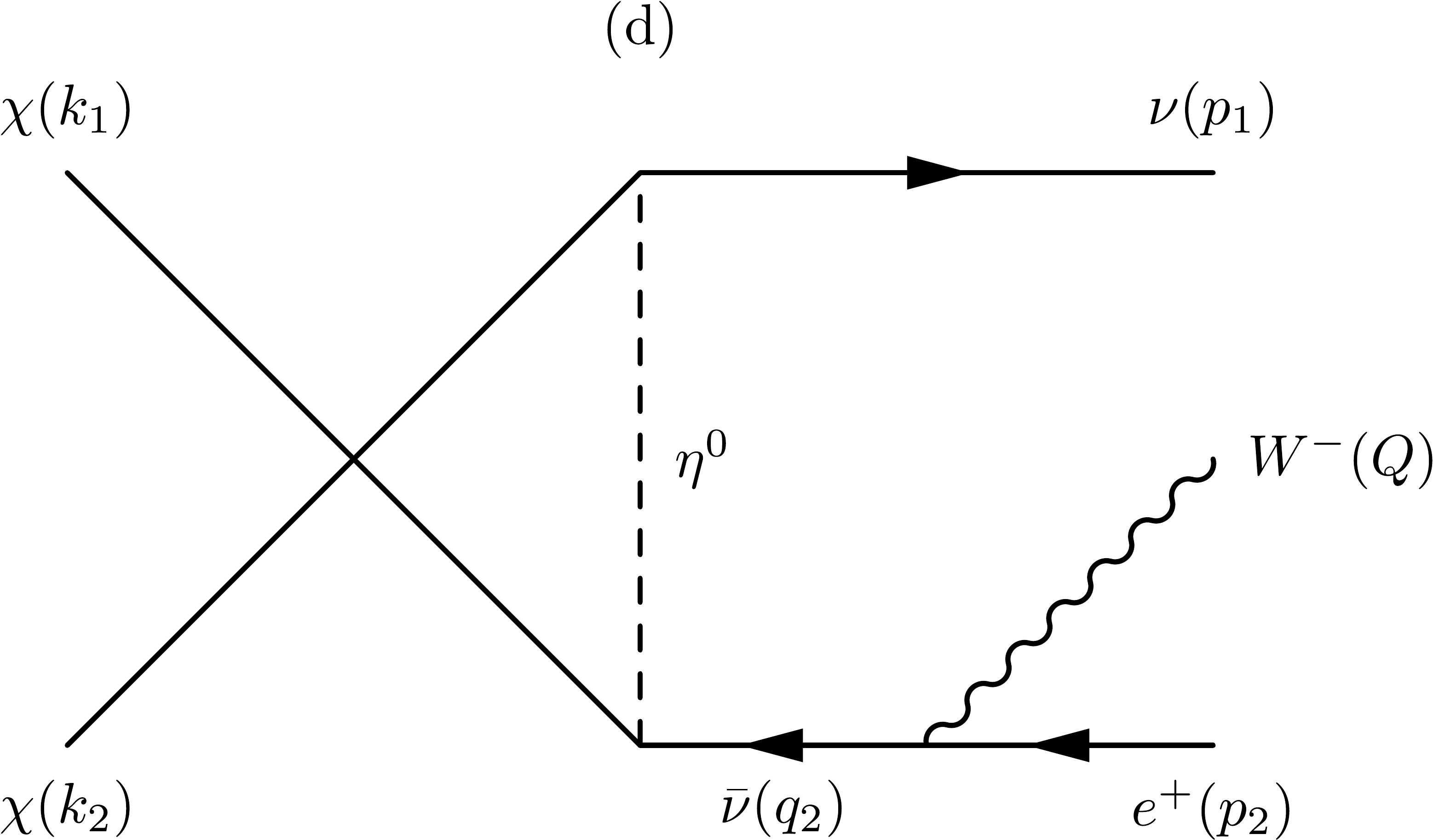}
$\;$
\includegraphics[height=0.10\textheight,width=0.32\columnwidth]{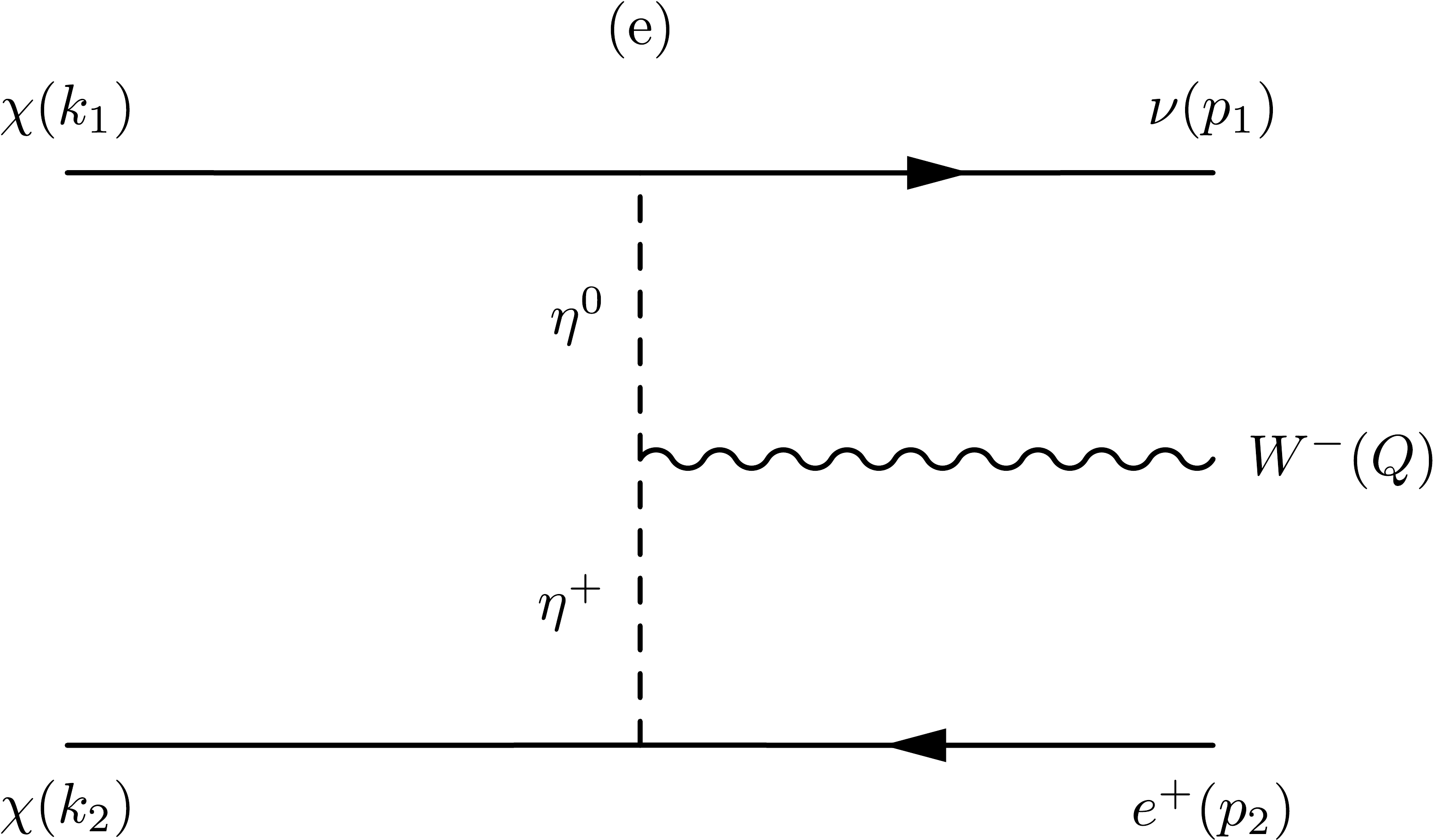}
$\;$
\includegraphics[height=0.10\textheight,width=0.32\columnwidth]{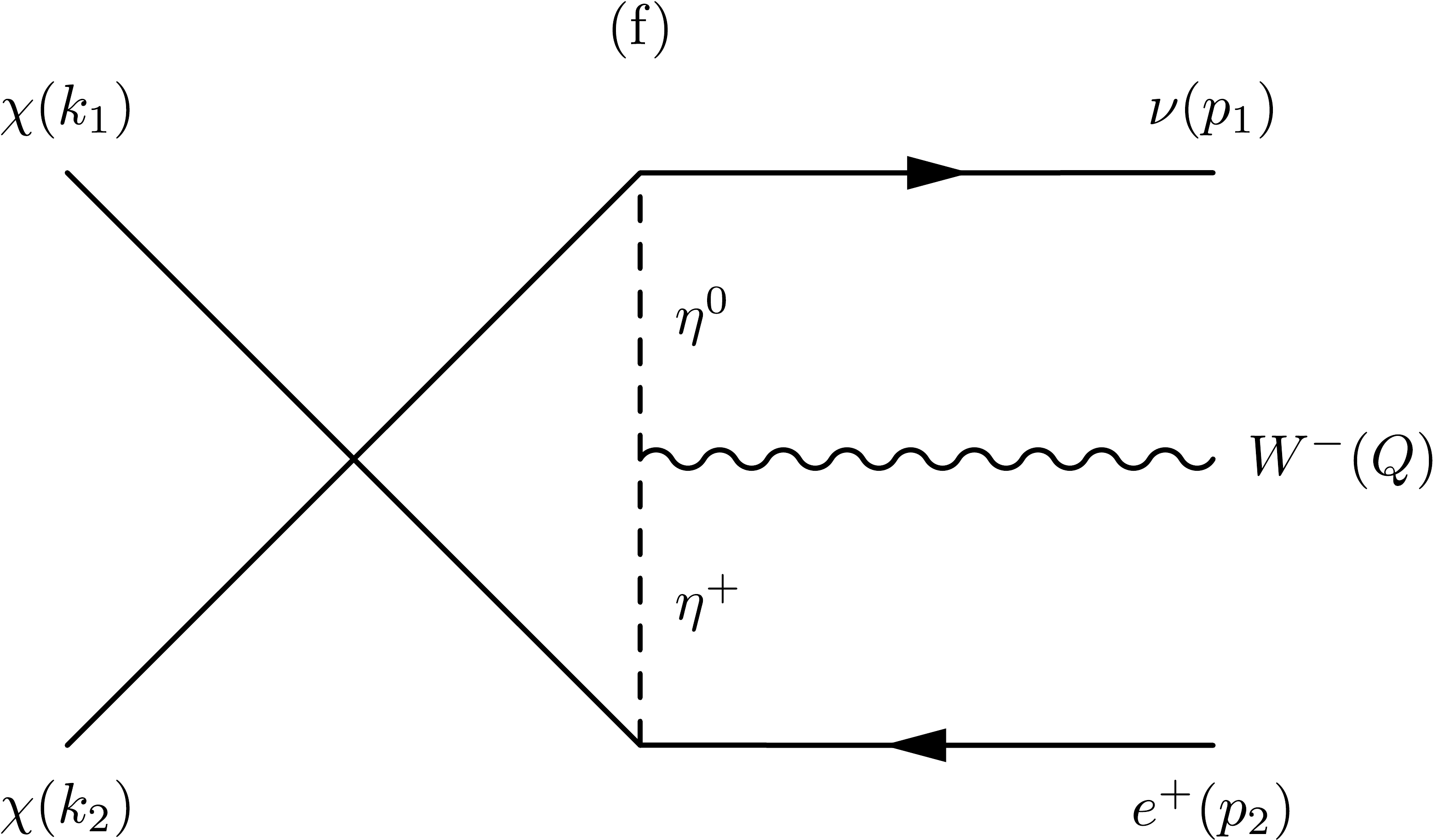}
\caption{
\label{fig:feyngraphs} 
The $t$- (left three) and $u$- (right three) channel Feynman diagrams for $\chi\chi\rightarrow e^+\nu W^-$.  
All fermion momenta in the diagrams flow with the arrow except $p_2$ and $q_2$, 
with $q_1=p_1+Q$, $q_2=p_2+Q$.}
\end{figure*}

Majorana-ness means that in the annihilation process to a fermion pair, $\chi\,\chi\rarr f\fbar$, 
the two initial state $\chi$'s contribute two amplitudes, the usual one and a crossed-$\chi$ diagram with a relative minus sign
(deducible from  Fig.~\rf{fig:feyngraphs} when the external gauge boson lines are removed).
The process is mediated by a $t$- and $u$-channel exchange of a virtual scalar particle which we label $\eta$.
For fermionic dark matter, Fierz transformations effect the natural projection of 
$2\rightarrow 2$ $t$- and $u$-exchange processes into partial waves. 
When the fermion currents (connected fermion lines) 
are Fierz rearranged into ``charge retention'' order (a $\chi$ line and a light fermion line),
the result is an axial vector coupling in the $s$-channel, plus corrections proportional to the $\chi$'s velocity, v.
This comes about because 
for each $t$-channel diagram (the first three shown in Fig.~\ref{fig:feyngraphs}),
there is an accompanying $u$-channel diagram (the last three shown in Fig.~\ref{fig:feyngraphs}),
obtained by interchanging the momentum and spin of the two Majorana fermions.
The relative sign between the $t$- and $u$-channel amplitudes is ($-1$) in accord with Fermi statistics.
Therefore, one obtains an elegant simplification: 
$V$ and $T$ couplings in the Fierzed bilinears of the $\chi$-current are 
zero to ${\cal O}({\rm v})$, and so ignorable.
Furthermore, the chiral nature of the fermion couplings offers no S or P terms after Fierzing.
Thus, after Fierzing, only the axial vector coupling of the $\chi$-current is significant.
It has been known for many years~\cite{Haim1983} that the spin/orbital~angular momentum of the 
$s$-channel axial-vector coupling requires a helicity flip of one of the produced fermions in the 
$L=0$ (``$s$-wave'') amplitude.
Thus, the $s$-wave amplitude is suppressed $\sim(m_f/M_\chi)$.
The $L=1$ (``$p$-wave'') amplitude does not require a helicity flip.
%, and so is not  $(m_f/M_\chi$-suppressed.
However, on general grounds, the $L^{th}$ partial wave amplitude
is suppressed as ${\rm v}^{L}$,
% ~\cite{note2},
and so the rate from the $p$-wave is suppressed as $\langle{\rm v}^2\rangle$.  
% In parametrized form, the rate can be written as $\langle{\rm v}\sigma\rangle = a +\langle{\rm v}^2\rangle\,b+\cdots$,
% where the constant $a$ comes from s-wave annihilation, while the
% velocity suppressed $\langle{\rm v}^2\rangle\,b$ term receives both s-wave and p-wave contributions.  

%%%%%%%%%%%%%%%%%%%%%%%%%%%%%%%%%%%%%%%%
%
% \begin{figure*}[t]
%\begin{figure}[t]
%\centering
% \includegraphics[height=0.18\textheight,width=0.32\columnwidth]{feyngraph-b}
% \includegraphics[height=0.18\textheight,width=0.32\columnwidth]{feyngraph-d}
% \includegraphics[height=0.18\textheight,width=0.32\columnwidth]{feyngraph-f}
% \caption{
% \label{fig:feyngraphs_bdf} 
% The $u$-channel Feynman diagrams. }
% \end{figure*} 
%%%%%%%%%%%%%%%%%%%%%%%%%%%%%%%%%%%%%%%%

Standard statistical mechanics 
simulations reveal that decoupling occurs at 
$T_{\rm dec}/M_\chi \sim \frac{1}{6}\langle {\rm v}^2\rangle\sim 1/20\ {\rm to\ }1/50$;
the resulting $\langle{\rm v}^2\rangle$ is therefore
$\sim 0.1\ {\rm to\ }0.3$, which implies only a mild $p$-wave suppression 
% is semi-relativistic.  Consequently, the $p$-wave contribution 
% in  $a+\langle{\rm v}^2\rangle\,b\sim 3\times 10^{-26}$cm$^3$/s 
%is velocity-suppressed by at most a factor of~0.1 
at DM decoupling in the Early Universe\@.
Today, v$\sim 300\ {\rm km/s}\sim 10^{-3}c$ 
in galactic halos, so the p-wave contribution is highly
suppressed by $\sim 10^{-6}$, and only the s-wave contribution is expected to be significant.  
However, as we have mentioned, in Majorana fermion DM models, 
the s-wave annihilation into a fermion pair $\chi\chi \rarr f\fbar$ 
is helicity suppressed by the factor $(m_f/M_\chi)^2$.
These two suppressions are quite general:
The helicity-suppression of the $s$-wave amplitude applies to all Majorana fermion $\chi\chi\rarr f\fbar$, 
while the velocity-suppressed $p$-wave applies to all DM~$\rarr f\fbar$, Majorana or otherwise.
% Alternatively, Majorana DM requires the contrivance of large ``boost factors'' 
% (e.g., sub-clustering of the DM toay, or ``Sommerfeld enhancement'' of annihilation) 
% to produce observable annihilation signals for indirect DM detection.  

It is becoming increasingly appreciated that if the light fermion pair is produced in association with a gauge boson,
then the spins of the final sate can match the $s$-wave angular momentum requirement without a helicity flip.
Thus, there is no $s$-wave mass-suppression factor for the $2\rarr 3$ process $\chi\chi\rarr f\fbar$+~gauge-boson.
This unsuppressed $s$-wave was first calculated in 1989 for the photon~bremsstrahlung reaction 
$\chi\chi\rarr  f\fbar\gamma$~\cite{gamma1,gamma2},
where the photon is radiated from one of the external particle legs
(final state radiation, FSR) or from a the virtual mediator particle $\eta$
(internal bremsstrahlung, IB).  
% Gauge invariance requires inclusion of both contributions.
On the face of it, the radiative rate is down by the usual QED coupling factor of
$\alpha/4\pi\sim 10^{-3}$.  However, and significantly, photon
bremsstrahlung can lift the helicity suppression of the $s$-wave
process, which more than compensates for the extra coupling factor.
(And if the dark matter annihilates to colored fermions, radiation of a gluon would also lift the helicity suppression.)

The importance of electroweak radiative corrections to dark matter
annihilation was recognized more recently.
Electroweak bremsstrahlung was investigated first in the context of cosmic 
rays~\cite{Berezinsky:2002hq,Kachelriess:2007aj,Bell:2008ey,Dent:2008qy,Ciafaloni:2010qr,Kachelriess:2009zy,Ciafaloni:2010ti};
the possibility of  $W/Z$~bremsstrahlung to lift initial-state velocity and 
final-state helicity suppressions was alluded to in~Refs.\cite{Bell:2008ey,Kachelriess:2009zy},
and the unsuppressed $s$-wave in the context of 
$\chi\chi\rarr f\fbar+W/Z$ was noted in~\cite{Bell:2010ei,Bell:2011if}.
Some signature channels for indirect detection of DM~$\chi$'s were calculated in Ref.~\cite{Bell:2011eu},
where the point was made that even if the $\chi\chi\rarr 2$ process is tuned to suppress direct $p\bar{p}$~production, 
the $2\rarr 3$ EW process will necessarily produce an antiproton signal via the production and hadronic decay of the $W$ and $Z$\@.
As a result, the experimental upper limit on the cosmic antiproton flux provides a meaningful constraint on Majorana DM~\cite{Bell:2011eu}.
The dominance of the helicity-unsuppressed $f\fbar W/Z$~channel has also been elaborated 
upon in Refs.~\cite{Ciafaloni:2011sa,Ciafaloni:2011gv}.

The diagrams contributing to EW bremsstrahlung are shown in Fig.~\ref{fig:feyngraphs}.  
There are a few important distinctions between electromagnetic (EM)
and electroweak (EW) bremsstrahlung. An obvious one is that EM bremsstrahlung
produces just photons, whereas EW bremsstrahlung and subsequent decay
of the $W$ or $Z$ leads to leptons (including positrons), hadrons (including antiprotons) 
and gamma rays, offering
correlated ``multi-messenger'' signals for indirect dark matter searches.
This is an important result for future DM indirect searches.
Another distinction between EW and EM~bremsstrahlung is that in the former 
the massive $W/Z$'s have a longitudinal mode not available to the photon.
The rate of  longitudinal $W_{\rm L}$  emission is proportional to the mass-squared splitting $M_{\eta^\pm}^2 -M_{\eta^0}^2$
of the two intermediate scalars in the IB~graph,
and may even exceed the radiation rate of the transverse~$W_{\rm T}$~\cite{Garny:2011cj}.

%%%%%%%%%%%%%%%%%%%%%%%%%%%%%%%%%%%%%
\begin{figure*}[!t]
\includegraphics[
%height=0.24\textheight,
width=0.39\textwidth]{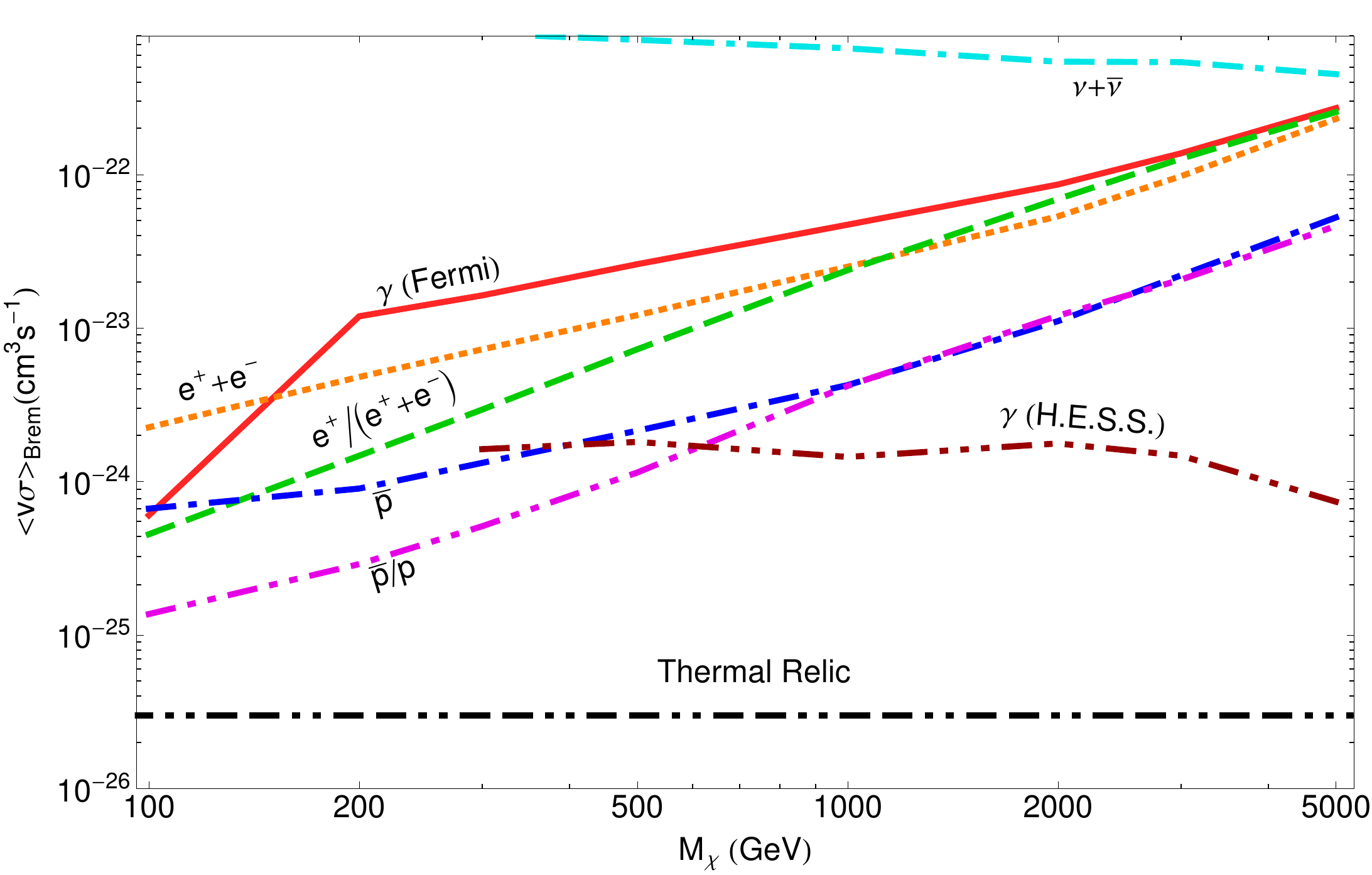}
% \\
\includegraphics[
height=0.17\textheight,
width=0.30\textwidth]{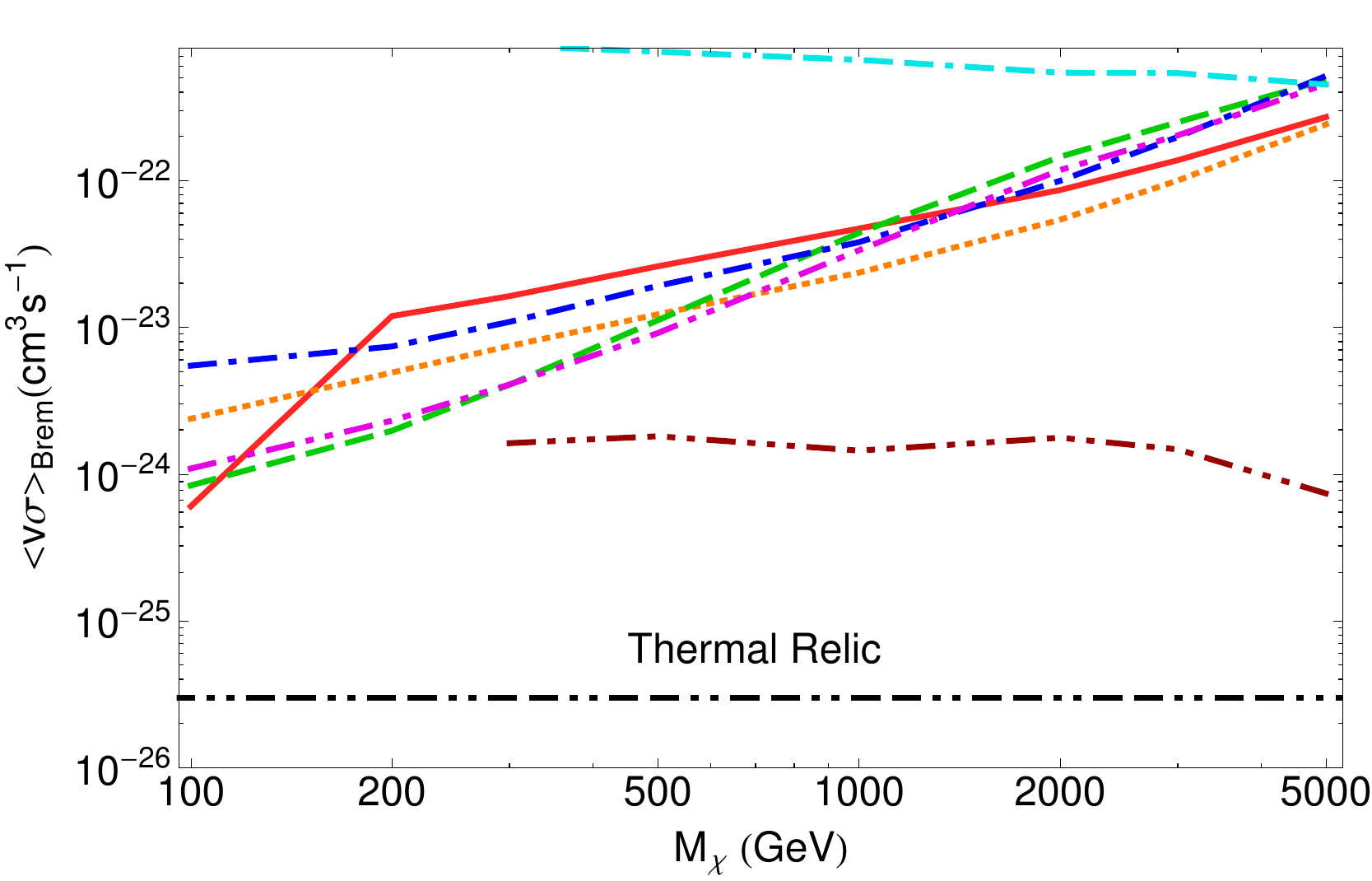} 
\includegraphics[
height=0.17\textheight,
width=0.30\textwidth]{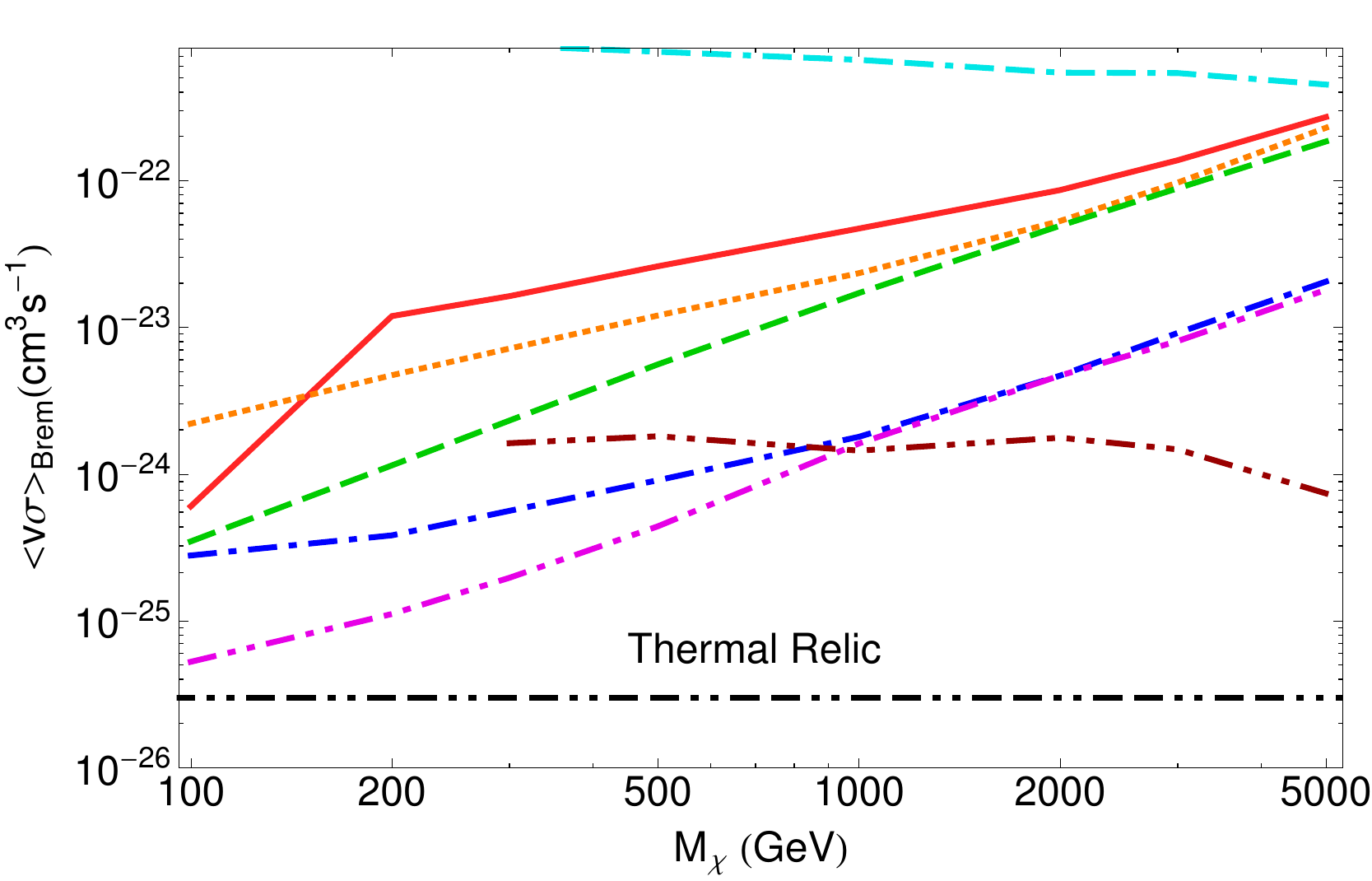}
\caption{Upper limits on $\langle v \sigma \rangle_{\rm Brem}$ using (left panel) 
  the `med' diffusion parameter set.  Shown are constraints based on
  the Fermi extragalactic background light (solid, red), 
  $e^+ +e^-$ flux (dots, orange),
  $e^+/(e^+ +e^-)$ ratio (dashes, green), 
  $\bar p$ flux (dot-dashes, blue), 
  $\bar p/p$ ratio (dot-dot-dashes, magenta),
  HESS $\gamma$'s (dot-dot-dot-dashes, maroon),
  and $\nu$'s (dot-dash-dashes, cyan).    
  Also shown is the
  expected cross section for thermal relic dark matter, $3\times
  10^{-26} \,\rm{cm}^3$/s.  % (dot-dot-dash-dashes, black).
  As for left panel, 
  but using the ``min'' (center) and ``max'' (right) diffusion parameter  sets.
\label{fig:limits-brem}}
\end{figure*} 
%%%%%%%%%%%%%%%%%%%%%%%%%%%%%%%%%%%%%

\section{Helicity and {\boldmath${\rm v}^2$} Suppressions Explored}
\label{sec:suppressions}
The discrete symmetries $C$, $P$, and $T$, and angular momentum conservation, 
place constraints on fermion pair ${\bar\chi}\chi$ states ~\cite{Bell:2008ey,Bell:2010ei},
as summarized in Table~\rf{table:LSJ}.  
A fermion pair can have total spin $S=1$ in the symmetric state or $S=0$ in the
antisymmetric  state.  The parity of the two-fermion state is
$P=(-)^{L+1}$, where $L$ is the orbital angular momentum of the pair.
% This parity formula holds for both Dirac and Majorana pairs.  The
% negative intrinsic parity of the pair, independent of the orbital
% parity $(-)^L$, is the same for Dirac and Majorana pairs for different
% reasons.  In the Dirac case, the $u$ and $v$ spinors (equivalently,
% the positive and negative energy states) are independent and have
% opposite parity corresponding to the $\pm 1$ eigenvalues of the parity
%operator $\gamma^0$.  Reinterpreting the two spinor types 
% (the positive and negative energy states) as particle and antiparticle
% then leads directly to opposite intrinsic parity for the
% particle-antiparticle pair.  In the Majorana case, the fermion has
% intrinsic parity $\pm i$, and so the two-particle state has intrinsic
% parity $(\pm i)^2=-1$.
% On general grounds, the $L^{\rm th}$~partial wave contribution to the
Since the annihilation rate is suppressed as v$^{2L}$, 
% where v is the relative velocity between the heavy, non-relativistic $\bar\chi \chi$ pair.   Thus 
only the $L=0$ partial wave gives an unsuppressed rate for DM in today's Universe.  
The general rule for charge-conjugation
(particle-antiparticle exchange) is $C=(-)^{L+S}$.
% The origin of this rule is as follows: 
% Under particle-antiparticle exchange, the spatial wave
% unction contributes $(-)^L$, and the spin wave function contributes
% (+1) if in the symmetric triplet $S=1$ state, and ($-1$) if in the
% antisymmetric $S=0$ singlet state, i.e., $(-)^{S+1}$.  In addition,
% there is an overall ($-1$) from anticommutation of the two
% particle-creation operators $b^\dag d^\dag$ for the Dirac case, and
% $b^\dag b^\dag$ for the Majorana case.
Two identical fermions comprise a Majorana pair,  
and so a Majorana pair is even under charge-conjugation, and from
$C=(-)^{L+S}$ one infers that $L$ and $S$ must be either both even,
or both odd for a Majorana pair.  

Consider the $L \le 2$ states.  In spectroscopic notation
$^{(2S+1)}L_J$ and spin-parity notation ($J^{PC}$), the vector $^3
S_1$~$(1^{--}$), $C$-odd axial vector $^1 P_1$~$(1^{+-})$, and
assorted $^3 D_J$~$(J^{--})$ states are all $C$-odd and therefore
disallowed for a Majorana pair.  
The pseudo-scalar $^1 S_0$~($0^{-+})$, scalar $^3
P_0$~($0^{++})$, axial vector $^3 P_1$~($1^{++})$, $C$-even tensor $^3
P_2$~($2^{++})$, and pseudo-tensor $^1 D_2$~($2^{-+}$) are all $C$-even
and therefore allowed.  In particular, the sole $L=0$~state, with no
v$^{2L}$ suppression, is the pseudo-scalar $^1 S_0$~($0^{-+})$.

At threshold, defined by $s=(2M_\chi)^2$ or
${\rm v}=0$, the orbital angular momentum $L$ is
necessarily zero.  With two identical Majorana fermions, the
two-particle wave function must be antisymmetric under particle
interchange.  Since $L=0$ at threshold, the $\chi\chi$ spatial wave
function is even, and the wave function must be antisymmetrized in its
spin.  The antisymmetric spin wave function is the $S=0$ state.  Thus,
the only contributing partial wave at threshold is the $^1 S_0$ state.
We have just seen that this is also the only state with no v$^{2L}$
suppression, so one may expect an unsuppressed Majorana annihilation
rate at threshold if and only if there is a $^1 S_0$ partial wave.

Finally, one may also invoke $CP$ invariance, the rule $CP=(-)^{S+1}$,
and the fact that $S=0$ and~1 are the only possibilities for a pair of spin 1/2 particles,
to deduce that total spin $S$ is conserved in any Dirac or Majorana annihilation to an $f\bar{f}$ final state.

%%%%%%%%%%%%%%%%%%%%%%%%%%%%%%%%%%%%%%%%%%%%%%%%%%%%%%%%%%%%%%%%%%%
%%%%%%%%%%%%%%%%%%%%%%%%%%%%%%%%%%%%%%%%%%%%%%%%%%%%%%%%%%%%%%%%%%%
\begin{table*}
%    \centering
        \begin{tabular}{|c|c|c|c|c|c|c|c|c|}
        \hline\hline
        $\mathbf{L}$ & $\mathbf{S}$ & $\mathbf{P=(-)^{L+1}}$ & $\mathbf{C=(-)^{L+S}}$ & $\mathbf{^{2S+1}L_J}$  & $\mathbf{J^{PC}}$ & \textbf{Name } 
        		& \textbf{Dirac Op} & $\mathbf{v^{2L}}$ \\ \hline\hline
        \multicolumn{9}{|c|}{\textbf{C-even states}} \\ \hline
        0   & 0   & $-$     & +     & $^{1}S_0$     & $0^{-+}$    & pseudo-scalar     &   $i\gamma_5$                      &  v$^0$   \\ \hline
        1   & 1   & +         & +    & $^{3}P_0$     & $0^{++}$    &   scalar                  &   $1$                     	                  &  v$^2$   \\ \hline
        1   & 1   & +         & +    & $^{3}P_1$     & $1^{++}$    & axial-vector          & $\gamma_5\gamma^k$      &  v$^2$    \\ \hline
        0   & 0   & $-$	& +	 & $^{1}S_0$     & $0^{-+}$     & 		            & $\gamma_5\gamma^0$	&  v$^0$   \\ \hline
        \multicolumn{9}{|c|}{\textbf{C-odd states (unavailable for Majorana pair)}} \\ \hline\hline
        0   & 1   & $-$    & $-$   & $^{3}S_1$      & $1^{--}$    & vector              &  $\gamma^k$     &  v$^0$          \\  \hline
        1   & 0   & +        & $-$   & $^{1}P_1$      & $1^{+-}$   &  	   	       & $\gamma^0$      &  v$^2$          \\  \hline
        1   & 0   & +        & $-$   & $^{1}P_1$      & $1^{+-}$   & tensor              & $\sigma^{jk}$      & v$^2$   	 \\  \hline
        0   & 1   & $-$    & $-$   & $^{3}S_1$      & $1^{--}$    &   		       & $\sigma^{0k}$    &  v$^0$           \\  \hline\hline
        \multicolumn{9}{|c|}{\textbf{Redundant C-even states unavailable for Majorana pair}} \\ \hline
        0   & 0   & $-$     & +     & $^{1}S_0$    & $0^{-+}$     & pseudo-tensor     & $\gamma_5\sigma^{jk}$      &  v$^0$    \\ \hline
        1   & 1   & +         & +    & $^{3}P_1$     & $1^{++}$    & 			            & $\gamma_5\sigma^{0k}$     &  v$^2$    \\ \hline\hline 
        \end{tabular}
    \caption{{\bf Decomposition of fermion bilinear currents into $s$-channel partial waves.} 
    Note that the relation 
    $\gamma_5\sigma^{\mu\nu}=\frac{i}{2}\epsilon^{\mu\nu}_{\ \ \alpha\beta}\sigma^{\alpha\beta}$, implies that 
    (i) the pseudo-tensor does not couple to Majorana fermions, and that 
    (ii) $P(\gamma_5\sigma^{\mu\nu})=P(\sigma^{\not\mu\not\nu})$, and 
    $C(\gamma_5\sigma^{\mu\nu})= - C(\sigma^{\not\mu\not\nu})$.
	    }
    \label{table:LSJ}
\end{table*}
%%%%%%%%%%%%%%%%%%%%%%%%%%%%%%%%%%%%%%%%%%%%%%%%%%%%%%%%%%%%%%%%%%%
%%%%%%%%%%%%%%%%%%%%%%%%%%%%%%%%%%%%%%%%%%%%%%%%%%%%%%%%%%%%%%%%%%%

What does this mean for a two-Majorana initial state which annihilates to a two-fermion final state?
The implications are best recognized after a 
Fierz transformation of the two mixed fermion bilinears to ``charge-retention'' order, i.e., 
to a $\chi$-bilinear and an f-bilinear~(many details may be found in an appendix of~\cite{Bell:2010ei}).
% As summarized in Table~\rf{table:LSJ}.
The Majorana pair couples only to the $C$-even basis fermion bilinears: the pseudo-scalar,
% $\Psibar i\gamma_5\Psi$~($0^{-+})$, 
the scalar,
% $\Psibar\Psi$~($0^{++}$), 
and the axial vector. 
% $\Psibar\gamma^\mu\gamma_5 \Psi$~($1^{++}$).  
%
% The vector $\Psibar\gamma^\mu\Psi$ and tensor $\Psibar\sigma^{\mu\nu}\Psi$
% bilinears are $C$-odd and therefore disallowed,
% while the pseudo-tensor bilinear $\Psibar i\gamma_5\sigma^{\mu\nu}\Psi
% =\frac{i}{2}\epsilon^{\mu]nu}_{\ \ \alpha\beta}\sigma^{\alpha\beta}$
% does not couple to a Majorana current because $\sigma^{\alpha\beta}$ does not.
If the Fierz'd bilinears contain a pseudo-scalar, there is no suppression of the rate.  
Otherwise, there is a v$^2$ rate suppression.  
If the Fierz'd bilinears contain an axial vector piece, 
it offers a $(m_f/M_\chi)^2$-suppressed $s$-wave contribution
(unless accompanied by a radiated $W$ or $Z$ or $\gamma$, as we shall show).\footnote
{ 
The axial-vector is an $L=1$ mode, and we have
seen that this mode elicits a v$^2$ suppression in the rate.  
However, the exchange particle is typically off-shell and so has an  $L=0$
timelike pseudo-scalar piece, with unsuppressed velocity-dependence,
but still with $(m_f/M_\chi)^2$ helicity-suppression.
}

% These deductions from partial wave analysis
% hold true for annihilation of Dirac or Majorana DM,
% while the deductions for $C$-even intial states are particular to the Majorana DM.

To address the question of which products of Fierzed currents are
suppressed and which are not, we set ${\rm v}^2$ to zero in the
$\chi$-current, and $m_f^2$ to zero in the fermion current.
In Table~\ref{table:bilinearlimits} we give the results
for the product of all standard Dirac bilinears.  
% Suppressed bilinears enter this table as zeroes.
(The derivation of these results is outlined in an appendix of~\cite{Bell:2010ei}.)
Read across rows of this table to discover that the only
unsuppressed $s$-channel products of bilinears for the $2\rarr 2$
process are those of the pseudo-scalar, vector, and tensor.  
For Majorana DM, the vector and
tensor bilinears are disallowed by charge-conjugation arguments 
%  again detailed in~\cite{Bell:2008ey}, 
and one is left with just the unsuppressed pseudo-scalar.
However, the couplings of chiral fermions to scalar $\eta$ contains no pseudoscalar (or scalar) 
in its Fierz transformation.

Reference to Table~\rf{table:bilinearlimits} reveals that the only component of the axial current 
which is non-vanishing in the $s$-wave (${\rm v}=0$ limit) is ${\bar\chi}\gamma_5\gamma^0\chi$.
However, there is no corresponding non-vanishing current $\bar\Psi\gamma_5\gamma^0\Psi$ or 
$\bar\Psi\gamma^0\Psi$ in the Table.  
Thus, the $s$-wave amplitude must be helicity suppressed.
Table~\rf{table:bilinearlimits} also reveals that a coupling of ${\bar\chi}\gamma_5\gamma^0\chi$ 
to $\bar\Psi\gamma_5\gamma^j\Psi$ or $\bar\Psi\gamma^j\Psi$ is helicity un-suppressed,
but requires a spin flip from parallel spinors in the initial state to antiparallel spinors in the final state.
Such a direct coupling would violate Lorentz invariance.
To the rescue comes  gauge boson emission,
which alters the fermion-pair spin state by one unit of helicity,
and couples ${\bar\chi}\gamma_5\gamma^0\chi$ 
to $\bar\Psi\gamma_5\gamma^j\Psi$ and $\bar\Psi\gamma^j\Psi$.
An un-suppressed $s$-wave amplitude will be the result. 

We now explain why the $t$- or $u$-channel scalar exchange
with opposite fermion chiralities at the vertices is so common.  It
follows from a single popular assumption, namely that the dark matter
is a gauge-singlet Majorana fermion.  As a consequence of this
assumption, annihilation to SM fermions, which are $SU(2)$ doublets or singlets,
requires either an $s$-channel singlet boson, or a $t$- or $u$-channel
singlet or doublet scalar that couples to $\chi$-$f$.  In the first
instance, there is no symmetry to forbid a new force between SM
fermions, a disfavored possibility.  In the second instance, unitarity
fixes the second vertex as the hermitian adjoint of the first.  
% Since the fermions of the SM are left-chiral doublets and right-chiral singlets, 
One gets chiral-opposites for the two vertices of the $t$-
or $u$-channel, and after a Fierz transformation, no unsuppressed pseudoscalar term.
So either the fermion bilinear is suppressed by $m_f$ in the $s$-wave or the $\chi$~bilinear 
is suppressed by v in the $p$-wave.

% Presented in Fig.~\ref{fig:rates}
% is a comparison of the $W$-strahlung rate to that for photon bremsstrahlung.  
% For high dark matter masses where the $W$ mass is negligible, 
% the two rates are identical except for the overall normalization, which is higher for $W$-strahlung 
% by the factor $1/(2\sin^2\theta_W)=2.17$ .  
% Another factor of two is gained for $W$-strahlung when the $W^+$ mode is added 
% to the $W^-$ mode shown in the figure.

%%%%%%%%%%%%%%%%%%%%%%%%%%%%%%%%%%%%%%%%%%%%%%%%%
%%%%%%%%%%%%%%%%%%%%%%%%%%%%%%%%%%%%%%%%%%%%%%%%%
\begin{table*} %[thbp]
%\begin{center}
\begin{tabular}{||c|c||c|c||c|c||}
\hline\hline
\multicolumn{2}{||c||}{s-channel bilinear $\Psibar\,\Gamma_D\,\Psi$} 
	& \multicolumn{2}{c||}{${\rm v}=0$ limit (projects out pure $s$-wave)} 
   	& \multicolumn{2}{c||}{${\rm M}=0$ limit (reveals helicity suppression)} \\ \cline{3-6}
\multicolumn{2}{||c||}{} & parallel spinors & antiparallel spinors & parallel spinors & antiparallel spinors \\  \hline\hline
scalar        & $\Psibar\,\Psi$                         
   & 0 & 0 & $ \sqrt{s}$ & 0 \\ \hline\hline
pseudo-scalar & $\Psibar\,i\gamma_5\,\Psi$               
   & $-2iM$ & 0 & $-i\sqrt{s}$ & 0   \\ \hline\hline
axial-vector  & $\Psibar\,\gamma_5\,\gamma^0\,\Psi$     
   & $2M$ & 0 & 0 & 0   \\ \cline{2-6}
              & $\Psibar\,\gamma_5\,\gamma^j\,\Psi$                                     
   & 0 & 0 & 0 & $\sqrt{s}\,(\pm\delta_{j1}-i\delta_{j2})$ \\ \hline\hline
vector        & $\Psibar\,\gamma^0\,\Psi$               
   & 0 & 0 & 0 & 0 \\ \cline{2-6}
              & $\Psibar\,\gamma^j\,\Psi$          
   & $\mp 2M\,\delta_{j3}$ & $-2M\,(\delta_{j1}\mp i\delta_{j2})$ & 0 & $-\sqrt{s}\,(\delta_{j1}\mp i
\delta_{j2})$ \\ \hline\hline
tensor        & $\Psibar\,\sigma^{0j}\,\Psi$        
   & $\mp 2iM\,\delta_{j3}$ & $-2iM\,(\delta_{j1}\pm\delta_{j2})$ & $-i\sqrt{s}\,\delta_{j3}$ & 0 \\ \cline
{2-6}
        & $\Psibar\,\sigma^{jk}\,\Psi$        
   & 0 & 0 & $\pm\sqrt{s}\,\delta_{j1}\delta_{k2}$ & 0 \\ \hline\hline
pseudo-tensor  & $\Psibar\,\gamma_5\,\sigma^{0j}\,\Psi$ 
   & 0 & 0 & $\pm i\sqrt{s}\,\delta_{j3}$ & 0 \\ \cline{2-6}
   & $\Psibar\,\gamma_5\,\sigma^{jk}\,\Psi$ 
   & $\mp 2M\,\delta_{j1}\delta_{k2}$ & $-2M\,(\delta_{j2}\delta_{k3}\mp i\delta_{j3}\delta_{k1})$ 
         & $-\sqrt{s}\,\delta_{j1}\delta_{k2}$ & 0 \\ \hline\hline
\end{tabular}
%\end{center}
\caption{
Extreme non-relativistic and extreme relativistic limits for s-channel fermion bilinears.
For a term with an initial-state DM bilinear and a final-state SM bilinear to remain unsuppressed,
the DM bilinear must have a nonzero entry in the appropriate cell of the ``${\rm v}=0$ limit'' columns,
and the SM bilinear must have a non-zero term in the appropriate cell of the  ``${\rm M}=0$ limit'' columns. 
% Otherwise, the term is suppressed. 
Recall that antiparallel spinors correspond to parallel particle spins 
(and antiparallel particle helicities for the $M=0$ current), and vice versa.
Amplitudes are shown for $\ubar\,\Gamma_D\,{\rm v} = [ \bar{\rm v}\,\Gamma_D\,u ]^*$.
The two-fold $\pm$ ambiguities reflect the two-fold spin assignments for parallel spins, and 
separately, for antiparallel spins.
}
\label{table:bilinearlimits}
\end{table*}
%%%%%%%%%%%%%%%%%%%%%%%%%%%%%%%%%%%%%%%%%%%%%%%%%%%
%%%%%%%%%%%%%%%%%%%%%%%%%%%%%%%%%%%%%%%%%%%%%%%%%%

%%%%%%%%%%%%%%%%%%%%%%%%%%%%%%%%%%%%%%%%%%%%%%%%%%%%%%%%%%%%%

\section{{\boldmath $W/Z$}-strahlung and Cosmic Signatures}
\label{sec:signatures}
%
% On the experimental astrophysics front, the PAMELA satellite has
% observed a sharp excess in the $e^+/(e^-$+$\,e^+)$ fraction at energies
% beyond approximately 10 GeV~\cite{PamelaPositrons}, without a
% corresponding excess in the $]bar{p}/p$ data~\cite{PamelaAntiprotons}, 
% while the Fermi satellite~\cite{Fermi1}, ground-based HESS~\cite{Aharonian:2009ah}, 
% and AMS onboard the ISS~\cite{Aguilar:2013qda}
% have reported excesses in the $(e^-$+$\,e^+)$ flux at energies up to order 1 TeV.  
Recent data from several experiments reveal an excess of an astrophysical positron flux at energies up to ${\cal O}$(TeV),
while no excess of  antiprotons is seen.
In fact, the observed $\pbar$ flux is well reproduced by standard astrophysical processes.
% ~\cite{Serpico:2011wg}.
%So constraints from antiprotons are likely to be even stronger than presented here.
While new astrophysical sources are thought to ultimately be the mechanism behind this excess~(see, e.g.,~\cite{Serpico:2011wg}),
DM annihilation in the Galactic halo has been advanced as an alternative explanation.
Many of the popular models proposed to explain the positron excess invoke Majorana DM.  

Ref.~\cite{Bell:2011eu} presented the spectra of
stable annihilation products produced via DM annihilation including $\gamma/W$/$Z$-bremsstrahlung.
After accounting for propagation through the Galaxy, we set upper bounds on the
annihilation cross section via a comparison with observational data.
Fig.~\ref{fig:limits-brem} collects our upper limits on the
 bremsstrahlung rate $\langle v \sigma\rangle_{\rm Brem}$.  
% Although $\mu=(M_\eta/M_\chi)^2=1.2$ was chosen for display, our
% conclusions remain valid for any value of $\mu\alt 6$ where the
% bremsstrahlung processes dominate the $2\rarr 2$~body processes.  

While our analysis techniques are conservative, there are large
astrophysical uncertainties in the propagation of charged particles
through galactic magnetic fields, and in the DM density profile
which probably contains substructure.  Consequently, our
constraints are illustrative, but not robust.
We assumed astrophysical propagation parameters~\cite{Cirelli:2008id} 
which are consistent
with a `median' $\pbar$ flux~\cite{Donato:2003}.  However, by assuming
alternate parameters, e.g. from the `max' or `min' $\pbar$ flux scenarios, our
results may be strengthened or weakened by up to an order of magnitude,
as shown in panels of Fig.~\ref{fig:limits-brem}.
Our conclusions hold in all cases considered except for the extreme ``min'' 
choice.
% of  the diffusion parameter set.
% where the $e^+/(e^++e^-)$ limits become comparable to those for $\bar{p}/p$ 
% and the $e^++e^-$ limits become comparable to those for gamma rays.

%%%%%%%%%%%%%%%%%%%%%%%%%%%%%%%%%%%%%
% \begin{figure}[ht]
% \includegraphics[height=0.15\textheight,width=0.17\textwidth]{results-brem.eps}
% \\
% \includegraphics[height=0.15\textheight,width=0.15\textwidth]{results-min.eps} 
% \includegraphics[height=0.15\textheight,width=0.15\textwidth]{results-max.eps}
% \caption{(Left panel) Upper limits on $\langle v \sigma \rangle_{\rm Brem}$ using
%  the `med' diffusion parameter set.  Shown are constraints based on
%  the Fermi extragalactic background light (solid, red), 
%  $e^++e^-$ flux (dots, orange),
%  $e^+/(e^++e^-)$ ratio (dashes, green), 
%  $\bar p$ flux (dot-dashes, blue), 
%  $\bar p/p$ ratio (dot-dot-dashes, magenta),
%  H.E.S.S. gamma rays (dot-dot-dot-dashes, maroon),
%  and neutrinos (dot-dash-dashes, cyan).    
%  Also shown for comparison is the
%  expected cross section for thermal relic dark matter, $3\times
%  10^{-26} \,\rm{cm}^3$/s.  % (dot-dot-dash-dashes, black).
%
%  (Middle and right panels) As for Fig.~\ref{fig:limits-brem}, 
%  but using the ``min'' (middle) and ``max'' (right) diffusion parameter  sets.
% \label{fig:limits-brem}}
% \end{figure} 
%%%%%%%%%%%%%%%%%%%%%%%%%%%%%%%%%%%%%

For the ``med'' parameter set, the constraint from the antiproton ratio is stronger than that from the
positron data by a factor of $\sim 5$.
% Nature provides a unique value for $\langle{\rm v}\sigma\rangle$.  
Therefore, if the observed positron fraction were
attributed to the bremsstrahlung process, then the same process would
overproduce antiprotons by about a factor of five,
and thereby preclude a sizable Majorana fermion 
DM contribution.

\section{Conclusions}
If DM is Majorana in nature, then its $2\rarr 2$~annihilation to SM 
fermions is suppressed due to helicity considerations.  However,
both electroweak and photon bremsstrahlung lift this suppression,
thereby becoming the dominant channels for DM annihilation
(EW exceeding EM if the DM mass exceeds $\sim 150$~GeV).

Unsuppressed production and subsequent decay of the emitted $W$ and $Z$ gauge bosons will produce fluxes of hadrons,
including $\pbar$'s, in addition to $e^-$'s, $e^+$'s, $\nu$'s,  and $\gamma$'s.
Importantly, we find that the null $\pbar$ data make it
difficult for helicity-suppressed Majorana DM annihilation to two fermions to source 
the reported cosmic $e^+$ excesses.
The obstacle is the copious production and subsequent hadronic decay of the EW gauge bosons, 
which leads to a significant antiproton flux.

%%%%%%%%%%%%%%%%%%%%%%%%%%%%%%%%%%%%%%%%%%%%%%%%
%% BACKMATTER
%%%%%%%%%%%%%%%%%%%%%%%%%%%%%%%%%%%%%%%%%%%%%%%%

\vspace*{0.5cm}
\footnotesize{{\bf Acknowledgment:}{
%I wish to acknowledge my collaborators on this $s$-wave/$p$-wave analysis of annihilation modes of 
%Majorana dark matter, Nicole Bell, James Dent, and Thomas Jacques, originally, and  
%Ahmad Galea and Lawrence Krauss as well, ultimately.
This work was partially supported by U.S. DOE award DE-FG05-85ER40226.
}}

\end{document}